\documentclass[aps,prl,twocolumn,superscriptaddress,reprint]{revtex4-2}
\pdfoutput=1
\usepackage{amsmath}
\usepackage{graphicx}
\usepackage{bbm}
\usepackage{amssymb}
\usepackage{enumitem}
\usepackage{booktabs}
\usepackage{verbatim} 
\usepackage{soul}
\usepackage{mathrsfs}
\usepackage[linktocpage=true,
colorlinks=true,
urlcolor=blue,
anchorcolor=blue,
citecolor=blue,
filecolor=blue,
linkcolor=blue,
menucolor=blue,
]{hyperref}
\usepackage[dvipsnames]{xcolor}

\newcommand{\ex}{\mathrm{e}}
\newcommand{\diff}{\mathrm{d}}
\newcommand{\dd}{\mathrm{d}}
\newcommand{\R}{\mathbb{R}}
\newcommand{\vol}{\mathrm{vol}}

\newcommand{\hook}{\mathbin{\rule[.2ex]{.4em}{.03em}\rule[.2ex]{.03em}{.9ex}}}

\newcommand{\tp}{\texttt{p}}

\usepackage{color}

\def\nn{\nonumber}
\newcommand{\ii}{\mathrm{i}}

\newcommand{\Z}{\mathbb{Z}}

\newcommand{\cA}{\mathcal{A}}
\newcommand{\cF}{\mathcal{F}}

\newcommand{\cK}{\mathcal{K}}

\newcommand{\cL}{\mathcal L}
\newcommand{\cR}{\mathcal R}
\newcommand{\cD}{\mathcal{D}}
\newcommand{\cS}{\mathcal{S}}
\newcommand{\cU}{\mathcal{U}}

\newcommand{\mf}[1]{\mathfrak{#1}}

\newcommand{\PhibarP}{{\Phi}^{(\text{anom})}}

\newcommand{\barM}{W}

\newcommand{\iiaSpinorPoly}{\hat{\Phi}}

\newcommand{\weight}{w}
\newcommand{\sigmaIIA}{y}

\newcommand{\gi}{{(\text{GI})}}
\newcommand{\cs}{{(\text{CS})}}

\newcommand{\cC}{\mathcal{C}}

\newcommand{\Sigmanew}{\mathscr{F}}
\newcommand{\Snew}{I}

\begin{document}

\title{Odd-Dimensional Localization in Supergravity}

\author{Pietro Benetti Genolini}
\affiliation{D\'epartement de Physique Th\'eorique, Universit\'e de Gen\`eve, 24 quai Ernest-Ansermet, 1211 Gen\`eve, Suisse}

\author{Florian Gaar}
\affiliation{Mathematical Institute, University of Oxford, Woodstock Road, Oxford, OX2 6GG, U.K.}

\author{Jerome P. Gauntlett}
\affiliation{Abdus Salam Centre for Theoretical Physics, Imperial College, Prince Consort Road, London, SW7 2AZ, U.K.}

\author{Jaeha Park}
\affiliation{Abdus Salam Centre for Theoretical Physics, Imperial College, Prince Consort Road, London, SW7 2AZ, U.K.}

\author{James Sparks}
\affiliation{Mathematical Institute, University of Oxford, Woodstock Road, Oxford, OX2 6GG, U.K.}

\begin{abstract}
\noindent 
We establish a localization principle for odd-dimensional supergravity theories containing Chern--Simons interactions. 
Our construction relies on combining an equivariant extension of the gauge-invariant part of the action with an equivariant completion of the Chern--Simons anomaly form $\PhibarP$ to form a relative equivariant cohomology class.  
This leads to a universal fixed  point formula for the on-shell action of supersymmetric solutions that is expressed in terms of
$\PhibarP$ and an equivariant Euler class. We illustrate the formalism in $D=11$ and $D=5$ supergravity, with applications including the supersymmetric Casimir energy and superconformal index of M5-branes, as well as black ring and black lens solutions.

\end{abstract}

\maketitle

\enlargethispage{0.25\baselineskip}

\section{Introduction}\label{sec:intro}

Supergravity theories in odd spacetime dimensions generically contain Chern--Simons couplings. 
These encode important topological and quantum aspects of the theory and play a central role in 
understanding various areas of string/M-theory including  anomalies, holography, black hole physics and 
compactifications. Unlike gauge-invariant Lagrangians, Chern--Simons terms are defined through 
generalized characteristic classes in one dimension higher. As a consequence, they present a significant obstacle to applying 
the localization techniques of \cite{BenettiGenolini:2023kxp}, which allow supersymmetric observables to be computed from global data without requiring an explicit supergravity solution.

Here we show that odd-dimensional supergravity theories admit a natural localization principle 
which involves an interesting interplay between relative equivariant classes and the Chern--Simons terms.
While localization methods have previously been applied in odd dimensions \cite{Cassani:2024kjn, Colombo:2025ihp, Cassani:2025iix, BenettiGenolini:2025icr, Colombo:2025yqy, Park:2025fon,Gaar:2026nqq,Cassani:2026teb}, 
our approach provides a unified framework that seems to apply universally. 

Consider a general supergravity theory in odd $D$ spacetime dimensions, with a $D$-form Lagrangian of the form
\begin{align}\label{splitlag}
\mathcal{L} = \mathcal{L}^{\gi} + \mathcal{L}^{\cs}\, ,
\end{align}
and Euclidean action 
$\Snew=(16\pi G)^{-1}\int_M \mathcal{L}$.
Here $\mathcal{L}^{\gi}$ is gauge-invariant, while  $\mathcal{L}^{\cs}$ is a gauge-dependent Chern--Simons term that, in general, is not a global form on $M$.
The Chern--Simons action is defined by 
choosing a manifold 
$\barM$ of dimension $D+1$ with boundary $\partial\barM=M$, 
extending $\mathcal{L}^{\cs}$ to $\barM$
with globally defined \emph{anomaly form} $P=\diff \mathcal{L}^{\cs}$ on $\barM$, and defining
\footnote{\label{globalqn}Key to this is the
quantization condition $\int_{N} P \in 2\pi \ii \mathbb{Z}$ (in Euclidean signature)
for any compact manifold $N$ without boundary, 
so that $\ex^{-S}$ is independent of the choice of $\barM$ and the
extension $P$ of the Chern--Simons form on $M$. 
This quantization condition is often related to the index of an operator,
 and  anomalies, 
 which in turn 
can necessarily require including higher-derivative Chern--Simons terms in the action (see e.g. \cite{Witten:1996md}). For the examples in this paper it is important that $W$ is a spin manifold.}
\begin{align}\label{anomformdef}
\int_{M}  \mathcal{L}^{\cs} \equiv \int_{\barM} P\, .
\end{align}

Supersymmetric solutions generically have a supersymmetric Killing vector $K$ on $M$, which, for odd $D$,
we assume to be nowhere zero. We assume $K$ extends to a vector field on $\barM$.
The anomaly form 
 $P$ is associated with a generalized characteristic 
class, and this in turn 
admits a natural equivariant completion to a polyform $\PhibarP$ on $\barM$, with top-form given by $P$.
Via this construction, it is automatic that $\diff_{K}\PhibarP=0$ where $\diff_{K}\equiv\diff - K\hook$ denotes 
the equivariant exterior derivative. 
Here we show that supersymmetry provides a natural polyform
$\Phi^{\gi}$, whose top-form of degree $D$ is the gauge-invariant part of the on-shell Lagrangian, 
$\mathcal{L}^{\gi}$, and crucially it satisfies 
\begin{align}\label{dPhi}
\diff_K \Phi^{\gi}
=-\PhibarP|_{M}\, .
\end{align} 

As we explain in the Supplementary Material (SM), this means that $(\Phi^{\gi},\PhibarP)$ 
form a relative equivariant cohomology class, and we may compute the on-shell action 
using the Berline--Vergne--Atiyah--Bott (BVAB) formula \cite{BV:1982, Atiyah:1984px}  on the manifold with boundary $\barM$. 
This leads to our main result for the on-shell action 
for closed manifolds $M$  
\begin{align}\label{main}
\Snew 
= \frac{1}{16\pi G} \int_{\Sigmanew} \frac{\PhibarP}{e_{K}(N\Sigmanew)}\, .
\end{align}
Here $\Sigmanew\equiv\{K=0\}\subset \barM$ is the fixed point set of 
$K$ in $\barM$, with normal bundle $N\Sigmanew$ and associated equivariant Euler class $e_{K}$. 
Remarkably, \eqref{main} shows that the on-shell supergravity action is computed via 
equivariant localization of the associated anomaly form for the theory. 
The right hand side depends only on certain global topological invariants 
and weights of the vector field at fixed points. Moreover, 
the result 
is independent of the choice of $\barM$ (see footnote \cite{Note1}) and one can choose any convenient $\barM$. 
A specific but useful construction arises when $M$ admits an additional  
Killing vector $\ell$ which generates a circle fibration. In this case we can define $\barM$ by filling in the circle fibres to obtain
a disc bundle, as described in the SM.

For $D=11$ supergravity and $D=5$ gauged supergravity, we explicitly
demonstrate \eqref{dPhi}, \eqref{main} at the level of two spacetime derivatives 
and discuss extensions to higher-derivative theories later, together with applications.

\section{\texorpdfstring{$D=11$}{D=11} supergravity}
We first consider $D=11$ supergravity, following \cite{Gauntlett:2002fz} but 
working instead in Euclidean signature.
After eliminating the Ricci scalar using Einstein's equations, the (partially) on-shell action for bosonic solutions is given by
$I= (16\pi G_{11})^{-1} \int_{M}\cL$, where the 11-form $\cL$ is
\begin{align}\label{onshellell}
    \cL
        &=\cL^{\gi}+\frac{\ii}{6}C\wedge G\wedge G\,.
\end{align}
The gauge-invariant part is $\cL^{\gi}= \frac{1}{3}G\wedge * G$ and 
$G=\dd C$ is the field strength for the three-form potential $C$.
The anomaly form, defined on $W$, is given by $P=\frac{\ii}{6}G\wedge G\wedge G$.

Focusing on supersymmetric solutions, and taking an analytic continuation of \cite{Gauntlett:2002fz}, we consider the following bilinears in
a Killing spinor $\epsilon$: 
\begin{align}\label{bilinears}
K^\flat \equiv \overline{\epsilon}\mskip1mu\Gamma_{(1)}\epsilon\, , \quad 
\Omega \equiv \overline{\epsilon}\mskip1mu\Gamma_{(2)}\epsilon\, , \quad 
\Sigma \equiv \overline{\epsilon}\mskip1mu\Gamma_{(5)}\epsilon\, ,
\end{align}
where $\Gamma_{(r)}\equiv \frac{1}{r!}\Gamma_{\mu_1\cdots\mu_r}
\diff x^{\mu_1}\wedge \cdots \wedge \diff x^{\mu_r}$ 
 and $\overline{\epsilon} \equiv \epsilon^T \cC$, with $\cC$ being the charge conjugation matrix. 
In Euclidean signature $\epsilon$ is a complex Dirac spinor, 
and the forms \eqref{bilinears} are in general complex. 
The Killing spinor equation implies  the vector $K$ dual to $K^\flat$ is Killing, 
with $\mathcal{L}_K G = \mathcal{L}_K\Omega=\mathcal{L}_K\Sigma=0$. 

As first noticed in \cite{Nekrasov:2021ked}, there is an equivariantly closed extension of the four-form $G$ on $M$,
\begin{align}\label{phigee}
    \Phi^{G}=G+\Omega\,,
\end{align}
with $\dd_K \Phi^{G}=0$.
Notice that there is no zero-form component of $ \Phi^{G}$. 
Next, we define the following gauge-invariant polyform, whose top form is $\cL^{\gi}$:
\begin{align}
    \Phi^{\gi}=\frac{1}{3}\Phi^G\wedge(* G+\ii\Sigma)\,.
\end{align}
Using supersymmetry and the equation of motion for $C$ one verifies that
$\Phi^{\gi}$ satisfies 
\begin{align}
    \diff_K\Phi^{\gi}=-\frac{\ii}{6}\Phi^{G}\wedge \Phi^G\wedge \Phi^G\,.
\end{align}
In the next subsection, for a specific $W$, relevant to type IIA supergravity,
 we describe an explicit extension of $\Phi^G$ to $W$ and hence
establish \eqref{dPhi}. 

\subsection{Localization and type IIA}\label{subsec:IIA}

We now assume we have an additional Killing vector $\ell$ on $M$, as described in the SM, with $\barM$ a 
spin disc bundle over
a regular Euclidean spacetime $B$ of type IIA supergravity.
We introduce an adapted coordinate $x^{10} \sim x^{10} + 2\pi\ell_s$ along an orbit of $\ell$, where $\ell_s$ is the string length,
and locally write the global angular one-form 
as $\alpha = \dd x^{10} + C_1$, where $C_1$ is the 
one-form potential of type IIA with field strength $F_2 = \diff C_1$.
Following the conventions of \cite{Becker:2006}, we locally decompose $C$ and $G$ on $M$ via
\begin{align}
	C&=(C_{3}-B_2\wedge C_1)+B_2\wedge \alpha\,,\nn\\
	G&={F}_4+H_3\wedge\alpha\,,
\end{align}
where $C_3$ and $B_2$ are type IIA potentials on $B$ and 
we have defined the type IIA field strengths $H_3$ and ${F}_4$ via
\begin{align}
	H_3&\equiv \diff B_2\,,\qquad
	{F}_4\equiv \diff C_3-H_3\wedge C_1\,.
\end{align}

Next we decompose the two-form bilinear $\Omega$ on $M$ as
\begin{align}
	\Omega = -\iiaSpinorPoly_2+ \widetilde{K}\wedge \alpha\,,
\end{align}
where $\iiaSpinorPoly_2$ and $\widetilde{K}$ are spinor bilinears of type IIA, as is the Killing vector $\xi$, which is the projection of $K$ to $B$ \footnote{For the Lorentzian analogue of this calculation the bilinears $\iiaSpinorPoly_0$ and $\iiaSpinorPoly_2$ are components of the IIA polyform $\iiaSpinorPoly=\ex^{-\phi}\epsilon_1\otimes \bar\epsilon_2$, where $\phi$ is the dilaton and $\epsilon_{1,2}$ are the IIA spinors, as discussed in \cite{Legramandi:2018qkr}}. They satisfy 
$\xi\hook \widetilde{K} =0$ and  $\xi\hook\iiaSpinorPoly_2 =  -\iiaSpinorPoly_0 \widetilde{K}$,
where $\iiaSpinorPoly_0$ is a IIA spinor bilinear. 
Note that the function $\iiaSpinorPoly_0$ satisfies $\iiaSpinorPoly_0= K\hook\alpha$ and
at a fixed point in $\Sigmanew\subset \barM$ the weight of $K$ on the disc fibres is $\ell_s^{-1}\iiaSpinorPoly_0$.

We define $\Phi^{H_3}=H_3+\widetilde{K}$. The polyform for $G$ on $M$ can then be decomposed as
\begin{align}
	\Phi^G = ({F}_4-\iiaSpinorPoly_2)+\Phi^{H_3}\wedge \alpha\,.
\end{align}
The equivariant closure of $\Phi^G$ implies $\dd_\xi\Phi^{H_3}=0$.

We assume there exists a function $\sigmaIIA$ on $B$ such that for $\sigma = B_2+\sigmaIIA$ we have $\diff_\xi\sigma=\Phi^{H_3}$, which requires $B_2$ to be in a gauge where $\mathcal{L}_\xi B_2 =0$. Then the general considerations in the SM imply that the push-down of $\Phi^G$ to the base $B$ is the equivariantly closed polyform $\Phi^G_B = \Phi^G -\diff_K(\alpha\wedge \sigma)$,
or more explicitly:
\begin{align}
	\Phi^G_B=[{F}_4-B_2\wedge F_2]+[\iiaSpinorPoly_0 B_2-\iiaSpinorPoly_2-\sigmaIIA F_2]+\iiaSpinorPoly_0 \sigmaIIA\,.
\end{align}
Note that the top-form is the four-form Page flux of IIA.

The extension of $\Phi^G$ on $M$ to $\Phi^G_W$
on $\barM$ is then 
\begin{align} \label{newphibar}
	\Phi^G_W = \Phi^G_B+\dd_{K} [r^2\alpha \wedge (B_2+\sigmaIIA)]\,,
\end{align}
and satisfies $\dd_K\Phi^G_W=0$.
Here $r\in[0,1]$ is the radial coordinate on the disc fibres with the boundary of $W$ located at $r=1$ and the base at $r=0$, so
that $\Phi^G_W|_{M}=\Phi^G$ and $\Phi^G_W|_{B}=\Phi^G_B$.
We highlight that the extension $\Phi^G_W$ has a zero-form component, in contrast to 
$\Phi^G$ in \eqref{phigee} which does not.

The $D=11$ (partially) on-shell action is given by
\begin{align}\label{deefiveosa}
(16\pi G_{11})I =\int_M \Phi^{\gi} + \int_{\barM} \PhibarP\,,
\end{align} 
where
$\PhibarP=\frac{\ii}{6} \Phi^G_W\wedge \Phi^G_W \wedge \Phi^G_W$.
The fixed point contribution to 
$I$ for supersymmetric solutions of $D=11$ supergravity, or equivalently the corresponding IIA solutions, can  
be evaluated using \eqref{main}, \eqref{Sfixedbase} with 
$\PhibarP|_B=\frac{\ii}{6}\Phi^{G}_B\wedge \Phi^G_B\wedge \Phi^G_B$.
For example, the contribution for isolated fixed points (nuts) is explicitly given by 
\begin{align}
I = \frac{2\pi\ell_s}{16\pi G_{11}}\sum_{\mathrm{nuts\,in\,}B} \frac{\ii}{6}\frac{(2\pi)^5}{b_1b_2b_3b_4b_5}(\iiaSpinorPoly_0)^2\sigmaIIA^3\,.
\end{align}
This precisely matches the result derived directly in IIA from equivariant localization (in a particular ensemble)~\cite{Couzens:2026xmi}.

\section{\texorpdfstring{$D=5$}{D=5} supergravity}
We consider $D=5$ gauged supergravity coupled to $n$ Abelian vector multiplets. The Lorentzian theory is constructed using
a very special real manifold of dimension $n$, specified by a symmetric tensor $C_{IJK}$, with $I,J,K = 1, \dots, n+1$, with real coordinates $Y^I$ satisfying $\mathcal{V}(Y)\equiv\frac{1}{6} C_{IJK} Y^I Y^J Y^K = 1$, and metric
$G_{IJ}=-\frac{1}{2}\partial_I\partial_J\log\mathcal{V}\mskip2mu|_{\mathcal{V}=1}$. 
There are $n+1$ gauge fields $\cA^I$ with curvatures $\cF^I \equiv \dd \cA^I$.
We also have  
$n+1$ real Fayet--Iliopoulos (FI) parameters $\zeta_I$ and a potential $V_{(5)}$.
Ungauged supergravity can be recovered by setting $\zeta_I=0$.

As in \cite{BenettiGenolini:2025icr} we consider the Euclidean theory where all fields are complex but the FI parameters $\zeta_I$ and $C_{IJK}$ are real. After eliminating the Ricci scalar, the on-shell action for bosonic solutions is 
$I = (16\pi G_{5})^{-1} \int_{M}\cL$,
where 
\begin{align}
	 \cL= \cL^{\gi} + \frac{\ii}{6}C_{IJK}\cA^I\wedge \cF^J\wedge \cF^K.
	\end{align}
Here the gauge-invariant part of the Lagrangian is given by $\cL^{\gi}= -\frac{2}{3}V_{(5)}\vol_5 + \frac{2}{3}G_{IJ}\cF^I\wedge * \cF^J $.
The anomaly form, defined on $W$, is $P=\frac{\ii}{6}C_{IJK}\cF^I\wedge \cF^J\wedge \cF^K$.

Supersymmetric solutions 
solve the Killing spinor equations, which are parametrized by two Dirac spinors $\chi$ and $\tilde{\chi}$. We define 
$D=5$ Killing spinor bilinears via 
\begin{equation}\label{dfiveksebiliniears}
	\cS \equiv \overline{\tilde{\chi}} \chi \, , \quad \ \cK^\flat \equiv \overline{\tilde{\chi}} {\Gamma}_{(1)} \chi \, , \quad \  \cU \equiv \ii \overline{\tilde{\chi}} {\Gamma}_{(2)} \chi \, ,
\end{equation}
which in general are complex \footnote{Reality conditions can be imposed as discussed in \cite{BenettiGenolini:2025icr}}.
These satisfy a number of algebraic and differential relations, which can be found in \cite{Gauntlett:2003fk,BenettiGenolini:2025icr}.
In particular,  the vector $\cK$ dual to $\cK^\flat$ is a Killing vector with $\cL_\cK\cF^I=\cL_\cK Y^I=0$ and 
$\cL_\cK\cS=\cL_\cK\cU=0$.

As noticed in \cite{BenettiGenolini:2025icr} there is an equivariantly closed extension of the field strengths:
\begin{align}\label{polygauge}
    \Phi^I=\cF^I + \Phi^I_0\,,\qquad \Phi^I_0\equiv \ii \cS Y^I\,,
\end{align}
with $\dd_\cK \Phi^I=0$. Next, we can define the following gauge-invariant polyform, 
$\Phi^{\gi}=\Phi^{\gi}_5+\Phi^{\gi}_3+\Phi_1^{\gi}$ with
\begin{align}
    \Phi_5^{\gi}&=\cL_5^{\gi}\,,
\nn\\
    \Phi_3^{\gi}&=\frac{2}{3}\ii\zeta_I Y^I *\cU-\ii Y_I \cF^I \wedge \cK^\flat+\frac{2}{3}G_{IJ}\Phi_0^I * \cF^J\,, \nn\\
    \Phi_1^{\gi}&=\cS\cK^\flat\,.
\end{align}
Using the results of  \cite{BenettiGenolini:2025icr} we then deduce the key result  
\begin{align}\label{dequalsfivekeyrel}
 \diff_\cK\Phi^{\gi} =- \frac{\ii}{6}C_{IJK}\Phi^I\wedge  \Phi^J\wedge \Phi^K \,.
\end{align}
In the next subsection we consider a specific $W$ and an explicit extension of $\Phi^I$ to $W$, hence establishing
\eqref{dPhi}.

\subsection{Localization and Kaluza-Klein reduction}
We now assume we have an additional Killing vector $\ell$, as described in the SM, with $\barM$ a disc bundle over
a regular $D=4$ space $B$. 
Locally we introduce an adapted coordinate $x^5\sim x^5+2\pi$ along the orbit of $\ell$, and write the global angular form as $\alpha = \diff x^5-A^0$, where $A^0$ is the Kaluza--Klein one-form potential on $B$ with $F^0=\diff A^0$.

The bilinear Killing vector $\cK$ trivially extends to $\barM$. We define
the function $\weight = \cK\hook \alpha$, which at the fixed points of ${\cK}$ on $W$ is the weight of
${\cK}$ in the disc direction.
Locally, we decompose $\cA^I$ and $\cF^I$ on $M$ via
\footnote{To aid comparison, we are using the same notation that was used to carry out a dimensional reduction from $M$ to $B$ in \cite{BenettiGenolini:2025icr}.}:
\begin{align}\label{KKansatz}
    \cA^I &= \check{A}^I+\check{z}_1^I\alpha\,,\nn\\
    \qquad  \cF^I &= (\check{F}^I-\check{z}^I_1F^0)+\diff\check{z}_1^I\wedge \alpha\,,
\end{align}
where $\check{A}^I$ is a local one-form with $\check{F}^I=\dd \check{A}^I$ and $\check{z}_1^I$ is a scalar on the base $B$.
The polyform for the field strength can then be decomposed as
\begin{align}
	\Phi^I = (\check{F}^I-\check{z}^I_1F^0)+(\check{\Phi}^I_0-\weight\check{z}^I_1)+\dd\check{z}_1^I\wedge \alpha\,,
\end{align}
where we defined $\check{\Phi}^I_0=\Phi^I_0+\weight\check{z}^I_1$. We then have $\ell\lrcorner \Phi^I=-\dd_\xi \sigma$ with $\sigma=\check{z}_1^I$. By the arguments in the SM, the push-down of the polyform $\Phi^I$ to the base $B$ is $\Phi^I_B = \Phi^I-\dd_\cK(\check{z}_1^I\alpha)$; more explicitly
\begin{align}
	\Phi^I_B = \check{F}^I +\check{\Phi}^I_0\,.
\end{align}
Then the extension of $\Phi^I$ to $\barM$ is given by 
\begin{align}
	\Phi^I_W=\Phi^I_B+\dd_{\cK}(r^2 \check{z}_1^I \alpha)\,,
\end{align}
where $r\in[0,1]$ is the radial coordinate on the disc fibre and satisfies 
$\dd_{{\cK}}\Phi^I_W=0$, 
$\Phi^I_W|_{M}=\Phi^I$ and $\Phi^I_W|_B = \Phi^I_B$.

The $D=5$ on-shell action is given by 
\begin{align}\label{deefiveosag}
(16\pi G_5)I =\int_M \Phi^{\gi} + \int_{\barM} \PhibarP\,,
\end{align} 
where 
$\PhibarP=\frac{\ii}{6} C_{IJK}\Phi^I_W\wedge \Phi^J_W \wedge \Phi^K_W$ and computed using \eqref{main}, 
\eqref{Sfixedbase} with 
$\PhibarP|_B=\frac{\ii}{6}C_{IJK}\Phi_B^I\wedge \Phi_B^J\wedge \Phi_B^K$. 
On $B$ the fixed points $\Sigmanew_\xi$ of $\xi$ are either points (nuts) or two-dimensional surfaces (bolts) and we find
\begin{align}\label{deefivenutsboltsexp}
    & (16\pi G_5)   I  =\ii \sum_{\mathrm{nuts}}  (2\pi)^3\frac{C_{IJK}}{6} \frac{\check\Phi^I_0\check\Phi^J_0\check\Phi_0^K}{b_1b_2\weight} +\\
    &\ii \sum_{\mathrm{bolts}} (2\pi)^2 \frac{C_{IJK}}{6}\frac{\check\Phi^I_0\check\Phi_0^J}{b\weight}\Big[\int 3\check F^K+\frac{\check\Phi^K_0}{\weight} \diff\alpha-\frac{2\pi \check\Phi^K_0}{b}c_1(\cL)\Big]\,,\nn
\end{align}
where 
$\check\Phi^I_0$ is constant on the bolts.

We can make a comparison with the results of \cite{BenettiGenolini:2025icr} (ignoring the boundary contributions discussed in \cite{BenettiGenolini:2025icr}). There 
$I$ was expressed in terms of the on-shell action $I_{(4)}$ of
a $D=4$ gauged supergravity theory plus an additional term involving the integral over $B$ of a four-form $\Lambda_4$, which is closed
but gauge-dependent. 
Using localization, the contribution to 
$I$ from $I_{(4)}$ was expressed in terms of fixed point data at nuts and bolts of $B$ and 
precisely agrees with \eqref{deefivenutsboltsexp} \footnote{Compare (2.68) and (2.71) of  \cite{BenettiGenolini:2025icr}, for example,
with \eqref{deefivenutsboltsexp} and set  $\Delta x^5=2\pi$ and identifying $\weight = (\Phi_0^0)^\text{there}$.}.
We thus have proven that the integral of $\Lambda_4$ over $B$ in \cite{BenettiGenolini:2025icr}, when treated correctly \footnote{A correct treatment of the integral of $\Lambda_4$ in the approach of \cite{BenettiGenolini:2025icr},
would require, in general, breaking $M_4$ into patches glued together with gauge transformations and summing the various contributions. Our result says that this procedure would exactly give zero.}, gives 
an exactly vanishing result. 

We can also place the main conclusions of \cite{Cassani:2026teb} at the two-derivative level on a firm footing. Invoking the analysis of \cite{Colombo:2025ihp,Colombo:2025yqy}, it was argued in \cite{Cassani:2026teb} that the fixed point contribution to the $D=5$ on-shell action is obtained by equivariant integration of the anomaly polynomial, precisely as we have just established. What was overlooked in \cite{Cassani:2026teb}, however, is that the on-shell action \eqref{deefiveosa} contains two distinct contributions and that it is only by virtue of the identity \eqref{dequalsfivekeyrel} 
(lifted to $W$) that the boundary terms cancel in the BVAB theorem (cf. \eqref{Sfixed}). With this key observation, the conclusions of \cite{Cassani:2026teb} follow, and, in particular, the fixed point contribution can be computed using six-dimensional toric geometry.

\subsection{Example: black ring and black lens}

We now consider $D=5$ supersymmetric solutions with $U(1)^3$ isometry, whose topology is specified by a toric fan \cite{Cassani:2025iix,Colombo:2025yqy}. To be concrete, associated with black hole solutions with
non-trivial horizon topology,
consider $M$ with fan vectors
\begin{align}\label{toric_5d_Lp1}
	&V^0 = (0,0,1) \,,\qquad\quad \ \, \, V^1 = (1,0,0) \,,\nn \\
	&V^2 = (0,\tp,1-\tp) \,,\qquad V^3 = (0,1,0) \,,
\end{align}
for some integer $\tp$.
A fan vector $V^a$ vanishes at a codimension two subspace $\cD_a$, which is a lens space $L(\texttt{p}_a, \texttt{q}_a)$, where $\texttt{p}_a=|(V^{a-1}, V^a, V^{a+1}) |$, and $\texttt{q}_a = (V^a,V^{a+1},w^a) \mod \texttt{p}_a$, with $w^a$ an arbitrary vector satisfying $(V^{a-1},V^a,w^a) = 1$. Here we have a ``horizon" $\cD_1 \cong L(\tp,1)$, a ``bubble", $\cD_2 \cong S^3$ with
$\cD_0\cong S^3$, $\cD_3\cong S^1\times S^2$.
The nomenclature comes from interpreting this closed $M$ as arising from an asymptotically flat or $AdS$ solution 
which has a horizon and a bubble. To regulate the action we have glued an appropriate 
reference background, effectively removing the vacuum contribution, as in \cite{BenettiGenolini:2025icr}. 
This fan gives rise to a smooth $D=5$ geometry for $\tp = 0$ and $\tp=2$, corresponding to an $S^2\times S^1$ black ring \cite{Elvang:2004rt,Gauntlett:2004qy} and an $L(2,1)$ black lens \cite{Kunduri:2014kja}, respectively. We continue with these cases, but keep $\tp$ general in the formulae below.
Explicit supergravity solutions have only been constructed for ungauged supergravity.

Choosing a particular $U(1)_\ell \subset U(1)^3$, given by $\ell = ( \ell_1 , \ell_2 , \ell_3)$,
where $\gcd(\ell_1,\ell_2,\ell_3) = 1$, we obtain a disc fibration $D\hookrightarrow W\to B$. In general
$B = M/U(1)_\ell$ is a toric orbifold: indeed the divisors $D_a$ that generate $H_2(B)$ are spindles $\mathbb{WCP}^1_{[d_a,d_{a+1}]}$, 
with $d_a = (V^{a-1},V^a,\ell)$. If $d_a = \pm1$ for all $a$, then $\barM$ will be free of orbifold singularities~\footnote{It is possible to generalize this to the case of orbifolds, but we leave that to future work.} and we can immediately use \eqref{deefivenutsboltsexp} provided that $W$ is spin.
For the given fan \eqref{toric_5d_Lp1} 
with $\tp=0$, these conditions are solved for $\ell_1=\pm 1$, $\ell_2=\pm 1$ and $\ell_3$ being an odd integer.
For $\tp =2$, they are solved for $\ell_1 = \pm 1$ and $\ell_2 = -\ell_3 = \pm 1$. 
We continue with these cases, but it is illuminating to keep 
$\ell$ general in the formulae below~\footnote{For $\tp =0$ with $\ell_1=\pm 1$, $\ell_2=\pm 1$ and $\ell_3$ being an even integer, the quotient $B$ is smooth, but $W$ does not admit a spin structure.  On the other hand, \cite{Cheng:2025ikd} require $W$ to be spin for the quantization argument \eqref{intcee}. Similarly, for $\tp=2$ with $\ell_1=\pm 1$, $\ell_2=\pm 1$ and $\ell_3=0$, $B$ is smooth, but $W$ does not admit a spin structure.}.

The supersymmetric Killing vector can be written in the toric basis as $\cK = (1, \varepsilon_1, \varepsilon_2)$
where $\varepsilon_{1,2}$ are associated with the angular velocities of the horizon.
The weights of ${\cK}$ at the fixed points on $W$, the poles of the  spindles labelled by $a$,  
are given by
\begin{align}\label{general_weights}
	b_1^a & = d_a^{-1} (\cK,\ell,V^a) \,, \qquad b_2^a = - \ d_a^{-1} (\cK,\ell,V^{a-1}) \,,\nonumber\\
	w^a & = -  d_a^{-1} (V^{a-1},V^a,\cK) \,.
\end{align}

The $D=5$ solution is further specified by the quantized magnetic fluxes $\mf{n}^I$ through a non-trivial two-cycle $\mathcal{C} \in H_2(M,\Z) = \Z$. One can show that there is 
 the following relation between 5d and 4d fluxes: 
\begin{align}\label{npp}
	\mf{n}^I \equiv \frac{1}{2\pi} \int_{\mathcal{C}} \cF^I & = \ell_2 \check{\mf{p}}^I_1 - \tp \, \ell_1 \check{\mf{p}}^I_{2} \,,
\end{align}
where the 4d fluxes $\check{\mf{p}}^I_a$ through $D_a$ on $B$ can be computed using BVAB:
\begin{align}\label{BVAB_p}
	\check{\mf{p}}^I_a \equiv \frac{1}{2\pi} \int_{D_a} \check{F}^I = \frac{\left. \check{\Phi}_0^I \right\vert_a - \left. \check{\Phi}_0^I \right\vert_{a+1}}{(\cK,\ell,V^a)} \,.
\end{align}
These fluxes are not independent due to homology relations:
for the fan \eqref{toric_5d_Lp1} and choice of $\ell$, we have  $\check{\mf{p}}_0^I = \frac{\ell_3}{\ell_1} \check{\mf{p}}_1^I + (\tp-1)\check{\mf{p}}_2^I$, $\check{\mf{p}}_3^I = \frac{\ell_2}{\ell_1} \check{\mf{p}}_1^I - \tp \,\check{\mf{p}}_2^I$.
That there is an extra independent flux in $B$, compared to a single $\mf{n}^I$ in $M$, is due to a redundancy in the decomposition \eqref{KKansatz}. We fix this ambiguity by setting $\check{\mf{p}}_2^I = 0$ (see \cite{BenettiGenolini:2025icr}), which implies $\check{\Phi}_0^I |_2 = \check{\Phi}_0^I |_3 \equiv \varphi^I$.
Given \eqref{npp}, \eqref{BVAB_p}, we may solve for the fixed point variables $\left.\check{\Phi}_0^I\right\vert_a$ in terms of $\mf{n}^I$ and $\varphi^I$.

Substituting into \eqref{deefivenutsboltsexp}, the on-shell action is 
\begin{align}\label{finaldeefiveblens}
	I &= - \frac{C_{IJK}}{6} \Bigg[ \frac{(\varphi^I -  \mf{n}^I\varepsilon_2)(\varphi^J - \mf{n}^J\varepsilon_2)(\varphi^K - \mf{n}^K\varepsilon_2 )}{\varepsilon_1 \varepsilon_2} \nonumber \\
	& \ \ - \frac{ (1-\tp)\varphi^I\varphi^J\varphi^K}{\varepsilon_2 ((1-\tp)\varepsilon_1 - \tp \, \varepsilon_2)} + \frac{\ell_3^2}{\ell_1 \ell_2} \mf{n}^I \mf{n}^J \mf{n}^K \Bigg]\frac{ \ii \pi^2}{2G_5}\,.
\end{align}
Notice that we are able to separate out the $\ell$-dependence, which fixes the choice of extension $\barM$.
This dependence of  
$I$ on $\barM$ can be resolved by considering higher-derivative Chern--Simons terms, as we discuss in the next section. The upshot is that we should drop the last term in \eqref{finaldeefiveblens} to get our final result for black rings with $\tp=0$ and black lenses with $\tp=2$, which
then agrees, to this order, with \cite{Cassani:2025iix,Colombo:2025yqy}, after a redefinition of the $\varphi^I$.

\section{Higher derivatives}
So far we have focused on classical supergravity theories with two derivatives. 
It is natural to generalize to theories with higher derivatives not least because the  
Chern--Simons terms themselves typically include higher-derivative terms which are essential for 
Eq. \eqref{anomformdef} to be well-defined. More generally, the higher-derivative terms will encode various perturbative quantum gravity effects. While the gauge-invariant part of the Lagrangian
$\mathcal{L}^\gi$ is, in general, a highly complicated object, the Chern--Simons part $\mathcal{L}^\cs$
is known exactly. We conjecture that for general theories, Eq. \eqref{dPhi} will continue to hold (possibly up to an additional equivariant exact piece). This will then lead to the on-shell action being localized as in~\eqref{main} but now with $\PhibarP$ the equivariant completion of the full anomaly form. 

\subsection{\texorpdfstring{$D=11$}{D=11} and the M5-brane theory}

For $D=11$ supergravity there is an additional Chern--Simons term proportional to $C\wedge X_8$, 
where the eight-form $X_8=\frac{1}{192}(P_1^2-4P_2)$ is defined in terms of Pontryagin forms $P_1,P_2$ of the Riemann curvature two-form
 \cite{Vafa:1995fj, Duff:1995wd}. If our conjecture above is valid, we can compute the on-shell action using \eqref{main}, but now with 
 \begin{align}
 \label{quantumanomaly}
	\PhibarP = \frac{\ii}{6}(\Phi^G)^3 + \ii(2\pi \ell_p)^6 \Phi^G\wedge\Phi^{X_8}\, ,
\end{align}
where $\Phi^G$ is as in \eqref{phigee} and
${\Phi}{}^{X_8}$ is defined by 
replacing the Riemann curvature two-form $\mathcal{R}^{ab}$ on $\barM$ by its covariant equivariant extension 
$\mathcal{R}^{ab}-\frac{1}{2}(\diff {K}{}^\flat)^{ab}$, with $a,b=1,\ldots,{12}$ being tangent space indices.  We have also
written $(2\pi)^8\ell_p^9=16\pi G_{11}$.

We can use this to compute the large-$N$ Cardy-like limit of the superconformal index for the $6d$ $(2,0)$ SCFT of type $A_{N-1}$ arising on M5-branes. This is holographically dual to complex, supersymmetry-preserving deformations of the $AdS_7$ black hole with topology $\R^2 \times S^5\times S^4$, and conformal boundary $S^1\times S^5\times S^4$ \cite{Bobev:2023bxl}. 
To regulate the action using background subtraction as in \cite{BenettiGenolini:2025icr,Park:2025fon}, we glue an $EAdS_7 \cong S^1\times \R^6$ factor on their common boundary, which leads to computing the action on $S^7\times S^4$ with $W=\R^8\times S^4$ obtained by filling the $S^7$.

The Killing vector $K$ rotates $\R^8$ with weights $\epsilon_I$, $I=0, \dots, 3$, and rotates $S^4\subset \R^2_1\oplus\R^2_2\oplus\R$ with weights $b_1$, $b_2$. There are two isolated fixed points corresponding to the origin of $\R^8$ over each of the poles of $S^4$.
The Killing spinor is uncharged under $K$ and so we can take
\begin{equation}
\label{eq:Mtheory_Weights_Constraint}
	b_1 + b_2 + \sum_{I=0}^3 \epsilon_I = 0 \, .
\end{equation}

As in the reduction from $D=11$ to type IIA, the extension of $\Phi^G$ to $W$ has
a zero-form piece, and we can compute the M5-brane charge $N$: 
\begin{equation}
\label{eq:Mtheory_N}
	N \equiv \frac{1}{(2\pi\ell_p)^3} \int_{S^4} G = \frac{1}{2\pi \ell_p^3} \frac{\Phi^G_0\rvert_N - \Phi^G_0 \rvert_S}{b_1b_2} \, ,
\end{equation}
where $N$ and $S$ label the poles of $S^4$. The antipodal map on $S^4$
reverses its orientation, and 
under parity $G\mapsto -G$; extending this symmetry to the equivariant completion fixes  $\Phi^G_0\rvert_N = -\Phi^G_0 \rvert_S$~\footnote{Alternatively one can deduce this using the  $\mf{so}(5)$-equivariant completion of the Bott--Cattaneo form for $S^4$.}. 
 We then use \eqref{main} to compute the localized supersymmetric M-theory action, including the one-loop correction \eqref{quantumanomaly}. 
After using \eqref{eq:Mtheory_N}, we obtain the key result~\footnote{If we had started with the non-closed $D=11$ solution with topology 
$\R^2 \times S^5\times S^4$ then we would have obtained the same result, provided that we regulate by turning all integrals into equivariant integrals. 
For more general classes of solutions, this approach will be available and can be taken to define a supersymmetric regularization.}:
\begin{align}
\label{eq:Mtheory_Action}
	I
	&= \frac{2\pi\ii}{\epsilon_0\epsilon_1\epsilon_2\epsilon_3} \left[ \frac{(b_1b_2)^2}{24} N^3 + (2\pi)^4 N \, \Phi^{X_8}_0 \rvert_{\text{fp}} \right] \, , 
\end{align}
where
\begin{equation}
\label{eq:Mtheory_Phi08}
	\Phi_0^{X_8}\rvert_{\text{fp}} = \frac{1}{192(2\pi)^4} \left[ \left( \sum_{i=1}^6 w_i^2 \right)^2 - 4 \sum_{i<j} w_i^2 w_j^2 \right] \, , 
\end{equation}
and $\{ w_i\} = \{ \epsilon_0, \dots, \epsilon_3, b_1, b_2\}$, with the constraint \eqref{eq:Mtheory_Weights_Constraint}.
After rescaling the weights to set $\epsilon_0 = 2\pi\ii$, the $O(N^3)$ term in \eqref{eq:Mtheory_Action} precisely matches the on-shell action of the supersymmetric complex deformation of the most general $D=7$ black holes of \cite{Bobev:2023bxl}. Our constraint \eqref{eq:Mtheory_Weights_Constraint} corresponds in $D=7$ gauged supergravity to requiring that the spinors are anti-periodic around the thermal circle, which is contractible in the black hole solution. In addition, \eqref{eq:Mtheory_Action} provides a precise gravitational computation for the $O(N)$ term.

We can also make a comparison with some field theory results. 
 First, still with $\epsilon_0 = 2\pi\ii$, \eqref{eq:Mtheory_Action} reproduces the Cardy-like limit of the superconformal index of the $6d$ $(2,0)$ SCFT of type $A_{N-1}$ up to $O(N^0)$, which can be obtained by combining the results of \cite{Nahmgoong:2019hko, Ohmori:2021dzb} (see also \cite{Chen:2026fpe} for a different derivation). In particular, the constraint \eqref{eq:Mtheory_Weights_Constraint} becomes the ``second-sheet'' constraint necessary in field theory.
Furthermore, our argument provides a first-principles gravity derivation of the prescription used in  \cite{Nahmgoong:2019hko, Ohmori:2021dzb}.
Our result \eqref{eq:Mtheory_Action} was obtained by an equivariant integration of the characteristic class \eqref{quantumanomaly} on the twelve-dimensional $W$ using \eqref{dPhi}, \eqref{main}. However, writing $TW= T\R^8\oplus TS^4$, where $TS^4$ encodes the $\mf{so}(5)$ R-symmetry gauging, performing first the integral over $S^4$ leads to the degree-eight characteristic class representing the anomaly polynomial of the interacting theory of $N$ M5-branes \cite{Harvey:1998bx}
\begin{equation}
\label{eq:Mtheory_AnomalyPolynomial}
    A_8 = N^3 \frac{p_2(TS^4)}{24} + N \, X_8 + O(N^0) \, .
\end{equation}
Thus, the remaining integral is the equivariant integration over $\R^8$, where now the equivariant integration is intended as a regulator of the integral over a non-compact space. But this is precisely the prescription used in \cite{Nahmgoong:2019hko, Ohmori:2021dzb} to compute the Cardy limit of the superconformal index of the $6d$ SCFT.

Second, from \eqref{eq:Mtheory_Action} we can extract the supersymmetric Casimir energy of the same theory \cite{Assel:2015nca,Cassani:2021fyv, ArabiArdehali:2021nsx}. We let $\epsilon_0 = 2\pi\ii/\beta$, so that $\beta$ represents the radius of the thermal circle, and then effectively take the definition 
\begin{equation}
\label{eq:Mtheory_Casimir_Definition}
	E_{\rm susy} = \lim_{\beta\to \infty} \frac{\partial I}{\partial \beta} \, ,
\end{equation}
which leads to
\begin{equation}
\label{eq:Mtheory_Casimir_FromGravity}
    E_{\rm susy} = \frac{1}{\epsilon_1\epsilon_2\epsilon_3} \left[ \frac{(b_1b_2)^2}{24} N^3 + (2\pi)^4 N \, \Phi_0^{X_8} \rvert_{\stackrel{ \rm fp\quad}{\epsilon_0 = 0}} \right] \, .
\end{equation}
In this limit the constraint \eqref{eq:Mtheory_Weights_Constraint} reduces to $b_1 + b_2 + \sum_{i=1}^3 \epsilon_i = 0$, and thus, \textit{mutatis mutandis}, our result matches the supersymmetric Casimir energy found in \cite{Bobev:2015kza}. Moreover, as above, we have provided a gravity argument explaining the conjectural relation between the supersymmetric Casimir energy and  equivariant integration of the anomaly polynomial put forward in \cite{Bobev:2015kza}. Indeed, by the same argument leading to \eqref{eq:Mtheory_AnomalyPolynomial}, the expression \eqref{eq:Mtheory_Casimir_FromGravity} is precisely the equivariant integral of the anomaly polynomial \eqref{eq:Mtheory_AnomalyPolynomial} on $\R^6$, subject to the constraint $b_1 + b_2 + \sum_{i=1}^3 \epsilon_i = 0$.

The agreement of both the Cardy limit and $E_{\rm susy}$ with independent field theory results provides a strong check of our conjecture that \eqref{dPhi} remains valid at least up to the eight-derivative level.

\subsection{\texorpdfstring{$D=5$}{D=5} and the black ring/lens}
For $D=5$ supergravity we can write the equivariant extension 
of the anomaly form on $W$ as 
\begin{align}
\PhibarP & = \frac{\ii}{6} C_{IJK} \Phi^I_W \wedge \Phi_W^J \wedge \Phi_W^K \nonumber\\
& \ \  + \frac{\ii}{48} C_{I} \Phi_W^I\wedge \mathrm{tr}(\Phi^{\cR}_W\wedge \Phi^{\cR}_W)\,,
\end{align}
where $\Phi^I_W$ is the equivariantly closed extension of $\cF^I$, as in the text, and $(\Phi^{\cR}_W)^{ab}=\cR^{ab}-\frac{1}{2}(\diff {K}{}^\flat)^{ab}$ is the equivariant extension of the curvature two-form $\cR^{ab}$ 
on $W$. For physical applications we assume \
\begin{align}\label{intcee}
	\frac{\pi}{4G_5} \left[ C_{IJK} \mf{n}^I \mf{n}^J \mf{n}^K + C_I \mf{n}^I \right] \in 6\Z\,,
\end{align} 
for $\mf{n}^I \in \Z$, which holds, for example, after
compactifying $D=11$ supergravity on a Calabi--Yau three-fold~\cite{Cheng:2025ikd}. 

Assuming our conjecture is true, the higher-derivative correction to the on-shell action \eqref{deefivenutsboltsexp} for isolated fixed points is given by
\begin{align}
	(16\pi G_5)\Delta I_{\rm HD} & = \ii \frac{C_{I}}{24} \sum_{{\rm nuts}} (2\pi)^3 \check{\Phi}_0^I \frac{b_1^2 + b_2^2 + w^2}{b_1 b_2 w} \,.
\end{align}

We can now apply this to the black ring and lens examples with $M$ given by the toric fan \eqref{toric_5d_Lp1}. We find that \eqref{finaldeefiveblens} should be corrected by the addition of 
\begin{align}\label{LensHDcorr}
	& \Delta I_{\rm HD} = -\frac{ \ii \pi^2}{2G_5}\frac{C_{I}}{24} \Bigg[ \frac{1+\varepsilon_1^2+\varepsilon_2^2}{\varepsilon_1 \varepsilon_2} (\varphi^I - \mf{n}^I \varepsilon_2) \nonumber \\
    & - \varphi^I \Bigg(\frac{((\tp-1)\varepsilon_1+\tp \varepsilon_2)^2+(\tp-1)^2+\varepsilon_2^2}{(\tp-1)\varepsilon_2((\tp-1) \varepsilon_1+\tp \, \varepsilon_2)} \Bigg) \nonumber \\
	& + \frac{1+ \ell_1^2+\ell_2^2+\ell_3^2}{ \ell_1 \ell_2} \mf{n}^I \Bigg] \,.
\end{align}
Given the toric fan \eqref{toric_5d_Lp1}, the extension $\barM$ is in general an orbifold. 
However, as mentioned above, to be smooth and spin for 
$\tp=2$ we take $\ell_1 = \pm 1$ and $\ell_2 = -\ell_3 = \pm 1$, 
while for $\tp=0$ we take $\ell_1 = \pm 1$, $\ell_2 =\pm 1$ and $\ell_3$ being an odd integer.
Choosing such $\ell$,
given the integrality property in
\eqref{intcee}, one can see that the $\ell$-dependent terms in $I + \Delta I_{\rm HD}$ can be dropped in the path integral modulo $2\pi\ii\Z$. The result for black rings and black lenses agrees with \cite{Cassani:2026teb}, where  $\varphi^I_{\text{here}} = (1-\tp) \varphi^I_{\text{there}} + \mf{n}^I \varepsilon_2$. 

\section{Discussion}
We have presented a new method for evaluating the on-shell action for supersymmetric solutions of
supergravity theories with Chern--Simons terms. The key ingredients are the supersymmetric Killing vector,  
equivariant completions of the anomaly form, as well as the gauge-invariant part of the Lagrangian that together satisfy
\eqref{dPhi}. For closed manifolds $M$, we then obtain the fixed point result~\eqref{main}. 

For many applications 
one is interested in manifolds $M$ with boundary, and
the BVAB theorem will add boundary contributions to \eqref{main}. In addition, these should be supplemented by the
boundary terms that are required to give a good variational principle and to regulate the action. 
In the holographic context, there are specific boundary terms that are needed to implement holographic renormalization
in a supersymmetric scheme. We conjecture that \eqref{main}, with the fixed point set in the interior of $M$, gives the exact
supersymmetric action with all other boundary terms cancelling, and it would be interesting to prove this \footnote{Such a computation
has been carried out for $D=4$ gauged supergravity coupled to vector multiplets in \cite{BenettiGenolini:2024lbj}.}.
In addition, for manifolds with boundary one can instead implement a background subtraction procedure to regulate the action
where we glue a reference geometry to $M$ rendering the whole space compact \cite{BenettiGenolini:2025icr}. In this case one can utilize 
\eqref{main} to obtain a finite result and, 
as we saw for $D=5$ black ring/lens solutions
both in the holographic and asymptotically flat setting, as well as $D=11$ $AdS_7$ black hole solutions, this gives physically 
sensible results for the supersymmetric action. 

In a companion paper \cite{MtheoryAiry} we have also shown how our formalism can be used to compute Airy functions for SCFTs arising on M2-branes. There are many additional applications and we plan to report on these soon.

\section*{Acknowledgments}
We thank Edoardo Colombo, Chris Couzens, Vasil Dimitrov, Alice L\"uscher and Luigi Tizzano for helpful discussions.
We thank the Galileo Galilei Institute for Theoretical Physics for its hospitality and support during the workshop ``Pathways to Quantum Black Holes: from Effective Theories to Exact Methods".
This work was supported in part by STFC grants ST/X000575/1 and
ST/X000761/1, and SNSF Ambizione grant PZ00P2\_208666.
JP is supported by a Dean's PhD studentship at Imperial College.
FG is supported by an STFC studentship.

\bibliography{biblio}{}

\appendix

\section{Supplementary material: }

\section{Relative localization}
Here we present some mathematical background on relative equivariant cohomology 
for manifolds with boundary, and discuss the BVAB formula \cite{BV:1982, Atiyah:1984px} in this setting. 

Consider a pair of manifolds 
$(M,\barM)$ where $M=\partial \barM$ is the boundary 
of $\barM$ and $M$ is taken to be closed (i.e. compact without boundary).
We assume there is a vector field $K$ on $\barM$, 
which restricts to a vector on $M$ (also denoted by $K$), and we assume $K$ is nowhere zero on $M$.
All vector fields are assumed to lie inside the Lie algebra of 
a compact torus action.
The equivariant exterior derivative acting on polyforms on $M$ is $\diff_K\equiv \diff - K\hook\mskip2mu$, 
and similarly on $\barM$. 
In this setting, a \emph{relative equivariant class} 
is represented by a pair of  
 $K$-invariant polyforms $(\Phi,\Psi_W)$ on 
$(M,\barM)$ satisfying 
\begin{align}\label{pair}
\diff_K \Psi_W=0\, , \quad \diff_K \Phi =-\Psi_W|_M\,.
\end{align}  
The first equation in \eqref{pair} says that
$\Psi_W$ defines an equivariant cohomology class on 
$\barM$, with the second saying that restricting to $M$ this class is trivialized by $\Phi$.

Consider the integral
\begin{align}\label{Sgeneral}
\Snew \equiv \int_M \Phi + \int_{\barM} \Psi_W\, ,
\end{align}
where, as usual, the integral of a polyform means one integrates the specific form with the relevant degree.
We may evaluate \eqref{Sgeneral} using localization: first extend 
$\Phi$ to a polyform $f\Phi$ on $\barM$, 
where $f$ is a smooth, $K$-invariant function that is
1 in a collar neighbourhood of the boundary $M$ and zero in 
the interior of $\barM$. We then have
\begin{align}\label{Sfixed}
\Snew & = \int_{\barM} \diff_{K}(f\Phi) + \Psi_W \nonumber\\
& =  
\int_\Sigmanew \frac{\Psi_W}{e_{K}(N\Sigmanew)} -
\int_M\frac{K^\flat}{\diff_K K^\flat}\wedge \left(\diff_K\Phi+\Psi_W|_M\right)\nonumber\\ 
& =  
\int_\Sigmanew \frac{\Psi_W}{e_{K}(N\Sigmanew)}\, .
\end{align}
The first equality uses Stokes' theorem.
The second equality uses the BVAB localization formula 
with boundary, where $K^\flat$ is the one-form dual 
to $K$ (using any $K$-invariant metric), and note $\diff_KK^\flat=
\diff K^\flat - \|K\|^2$ has nowhere vanishing zero-form part so is invertible.  The last equality follows from  \eqref{pair}: the BVAB boundary term vanishes for a relative equivariant class.

The last line of \eqref{Sfixed} is the fixed point result: 
$\Sigmanew\equiv\{K=0\}\subset \barM$ is the fixed point set of 
$K$ in $\barM$, with normal bundle $N\Sigmanew$ and associated equivariant Euler class $e_{K}$, where note we have used $f|_\Sigmanew=0$. For generic action of $K$, 
 $N\Sigmanew=\oplus_{i=1}^r L_i$ splits as a sum of complex line bundles
$L_i$ where we can write $K=\sum_{i=1}^r b_i \partial_{\varphi_i}$, 
with $\partial_{\varphi_i}$ rotating $L_i$ with weight 1. Then 
\begin{align}\label{eqclasSM}
e_{K}(N\Sigmanew) = \wedge_{i=1}^r\Big[ \frac{b_i}{2\pi} + c_1(L_i)\Big]\, ,
\end{align} with $c_1(L_i)$ the first Chern class and $b_i\neq 0$ for all $i=1,\ldots,r$ ensures this is invertible.

\section{Disc fillings}

A specific example of the above construction arises as follows. 
We assume that in addition to $K$ we have another nowhere zero vector field 
$\ell$ on $M$, under which all forms are also invariant, with $[K,\ell]=0$, and which 
generates a circle fibration, 
\begin{align}
S^1_\ell\hookrightarrow M\to B\,,
\end{align}
with $B$ the base manifold \footnote{Generalizations to locally free actions where $B$ is an orbifold are also possible.}
and $\pi:M\to B$.
Note  
the fixed point set of the Killing vector $\xi=\pi_*(K)$ in $B$ is the projection of the subset of $M$ where $K$ and $\ell$ are aligned.
We let $\alpha$ be the global angular one-form for the circle fibration with $\ell\hook \alpha =1$, $\ell\hook \diff\alpha=0$. 
We then choose  $\barM$ to be the disc fibration
over $B$ which is obtained by filling in the circle fibres of $M$ 
\begin{align}
D\hookrightarrow \barM\to B\,.
\end{align}
We may choose a radial coordinate $r\in[0,1]$ on the disc fibres, with 
$M=\partial \barM=\{r=1\}$ and $B=\{r=0\}$. 

In this  setting, with a further assumption below, there is a canonical way to 
lift an equivariantly closed form $\Psi$ on $M$ to an equivariantly closed form $\Psi_W$ on $\barM$, 
and also push it down to an equivariantly closed form $\Psi_B$ on $B$. 
Since $\mathcal{L}_\ell = \diff_K\circ \ell\hook + \ell\hook \circ \mskip3mu \diff_K$ and 
by assumption $\diff_K\Psi=0=\mathcal{L}_\ell\Psi$, it follows that 
$\diff_K (\ell\hook \Psi)=0$. We solve this as $\ell\hook \Psi = -\diff_K \sigma$, 
where the polyform $\sigma$ on $M$ is only a global polyform if we further assume
the equivariant cohomology class of $\ell\hook\Psi$ is zero.
Then one can check that 
\begin{align}
\Psi_B &\equiv \Psi -\diff_K(\alpha\wedge \sigma)\, ,\nn\\
\Psi_W &\equiv \Psi_B + \diff_{K}(r^2\alpha\wedge\sigma)\, ,
\end{align}
define smooth polyforms on $B$, $\barM$, satisfying 
$\diff_\xi \Psi_B=0=\diff_{K}\Psi_W$, $\Psi_W|_M=\Psi$, $\Psi_W|_B=\Psi_B$. 
In particular note that $\ell\hook \Psi_B=0$ and $r^2\alpha$ is a smooth one-form on $\barM$ that vanishes on $B$, where $r=0$. Moreover, 
$\diff_K\alpha=\diff\alpha - \weight$ is a representative 
of $-2\pi$ times the first Chern class of $L(D)$, the disc 
 bundle over $B$, where $\weight\equiv K\hook\alpha$.  The function $\weight$ is precisely the weight of $K$ on the 
disc fibres over a fixed point in $\Sigmanew\subset \barM$.  

The normal bundle of $\Sigmanew$ splits as $N\Sigmanew =L(D)\oplus N\Sigmanew_\xi$ and the equivariant Euler class then decomposes as
\begin{align}
    e_{K}(N\Sigmanew) = \Big(\frac{\weight}{2\pi}-\dd\alpha\Big)\wedge e_\xi(N\Sigmanew_\xi)\,.
\end{align}
There is an expression analogous to \eqref{eqclasSM} for $e_\xi(N\Sigmanew_\xi)$.
Putting this altogether, the integral \eqref{Sfixed} can be written
\begin{align}\label{Sfixedbase}
\Snew=
\int_\Sigmanew \frac{\Psi_B}{\big(\frac{\weight}{2\pi}-\dd\alpha\big)\wedge e_\xi(N\Sigmanew_\xi)}\, ,
\end{align}
where $\Psi_W=\Psi_B+\dd_{K}(r^2\alpha\wedge\sigma)$, with $\ell\hook \Psi = -\dd_K \sigma$.

Finally, we highlight that we are interested in $W$ that are spin, which requires  $c_1(L(D))=w_2(B)$ mod $2$, with $w_2$ the second Stiefel--Whitney class.

\end{document}